\documentclass[aps,twocolumn,floats,prd,nofootinbib,superscriptaddress,showkeys,10pt]{revtex4-2}
\usepackage{graphicx}

\usepackage{graphicx}
\usepackage{bm}
\usepackage{placeins}
\usepackage{xspace}
\usepackage{bm,amsmath,amssymb,amsfonts,gensymb}
\usepackage{comment}
\usepackage{hyperref}
\hypersetup{
    colorlinks=true,
    linkcolor=blue,
    citecolor=blue,
    filecolor=blue,      
    urlcolor=blue,
}

\usepackage{color}

\usepackage[ukrainian,english]{babel}
\usepackage{verbatim}

\begin{document}
	
\title{Penrose hypothesis and instability of naked singularities in static spherically symmetric systems with scalar fields
}

\author{~A.~V.~Tugay}
\affiliation{Taras Shevchenko National University of Kyiv,  Kyiv 01601, Ukraine}
\author{~V.~I.~Zhdanov}
\affiliation{Taras Shevchenko National University of Kyiv,  Kyiv 01601, Ukraine}
\affiliation{Igor Sikorsky Kyiv Polytechnic Institute, Kyiv  03056, Ukraine}
\author{~Yu.~V.~Taistra}
\affiliation{Pidstryhach Institute for Applied Problems of Mechanics and Mathematics NAS of Ukraine, Lviv 79060, Ukraine}
\affiliation{Lviv Polytechnic National University, Lviv 79013, Ukraine}

\begin{abstract}
General relativistic static spherically symmetric (SSS) asymptotically flat configurations with  scalar fields  typically contain naked singularities at the center. We consider minimally coupled scalar fields  with power-law potentials leading to the Coulomb asymptotic  of the field $\phi(r)\approx Q/r$ for large values of the radial variable $r$. The configurations are uniquely defined by total mass and a $Q$-parameter  characterizing the strength of the scalar field at spatial infinity. The focus is on the linear stability against radial (monopole) perturbations of the SSS configurations satisfying conditions of asymptotic flatness. Our numerical investigations show the existence of divergent modes of small perturbations against the static background, at least for sufficiently  small values of $Q$. This means instability of the  configurations, confirming the well-known Penrose conjecture about the nonexistence of naked singularities - in this particular case.  On the other hand, we have not found divergent modes of linear radial perturbations for sufficiently large $Q$. 
\end{abstract}


\maketitle{Keywords: cosmic censorship,  scalar fields, naked singularities}

\section{Introduction} \label{Introduction}
The evolution and structure  of space-time singularities  present an unsolved problem  of classical General Relativity (GR). There is no general answer to the question, except a few exact and/or approximate special solutions,  how the singularities move after they are formed, can they evolve into black holes, can they disappear etc.  The existing physical folklore connects these problems with an  incompleteness of GR, which can probably be overcome by some modification, possibly by unification with a quantum field theory. On the other hand, there are attempts to solve at least some of the above problems without going beyond GR. 

There are two main types of the space-time singularities in  GR  and its modifications. The naked singularity (NS)  is, by definition,  a space-time singularity that can be viewed by a distant observer and   influence  its future. The other type is represented by singularities inside black holes that are hidden under the event horizons.
According to the  cosmic censorship conjecture by Roger Penrose \cite{Penrose1965, Penrose2002}, the NSs should be absent in the real universe; all   the space-time singularities  must be hidden under the event horizon from external observers. Although  this looks like a wishful thinking, this hypothesis is very popular. The motivation for this claim is that in presence of a naked singularity the evolution governed by Einstein equations may be nonunique  and the theory of gravity may lose its predictive power. But there exist prescriptions  \cite{Wald1980,Ishibashi-Hosoya1999,Ishibashi-Wald2003,Gibbons2005}  that guarantee    a well defined
dynamic at least for some systems having  NSs.   On the other hand, there are counterexamples to the  Penrose hypothesis that show how the NS can be formed \cite{Christodoulou1984,Ori1987,Joshi1993,Joshi_2013,Ong2020,Roberts1989}; although there may be  questions about feasibility of the physical conditions corresponding to these counterexamples.

If NS can be formed somehow, then the question of its stability arises.  An instability of SSS isolated gravitational system could be an  argument in favor the Penrose hypothesis. However, numerous publications  on the stability of spherically symmetric objects in various theories of gravity (see, e.g.,  \cite{Bronnikov-Zhidenko2011,Bronnikov2024}), following the first works on this topic  \cite{Bronnikov1979,Jetzer1992,Clayton1998}, show that the answer to the  question of stability may be different for different models with NSs. This answer  may be different even within the same model, but for different configuration parameters.

The first step in this direction is due to the linearized stability against small perturbations \cite{Bronnikov-Zhidenko2011,Bronnikov2024,Bronnikov1979,Jetzer1992,Clayton1998}.  We follow this way in case of 
SSS configurations with massless SFs described by the power-law potentials. Note that the stability of  SSS configurations with these potentials has been studied in \cite{StSavZh2024}; it has been shown that SSS configurations are linearly stable against axial perturbations. 
However, the monopole (radial) perturbations, which just might be suspected of causing instabilities \cite{Bronnikov2024}, have not been investigated. Here, we fill this gap and provide a consistent consideration for a set of the background configuration parameters, keeping in mind that to prove instability we need at least one divergent (as a function of time) solution: if such a solution exists, there is no need to study  other types of perturbations.  

In Section \ref{Sect1_general} we present basic relations describing the background SSS gravitational configuration with SF. Section \ref{Section2_perturbations}  deals with radial perturbations. Here the one-dimensional master equation for the  monopole (radial) perturbations is derived and  boundary conditions for a corresponding eigenvalue problem are formulated. We describe the shooting method used and present the results of the numerical simulations  depending upon the configuration parameters, in particular, upon the scalar charge $Q$  characterizing the strength of the scalar field at spatial infinity. The results are discussed in Section \ref{Section-conclusions}.

\section{Spherically symmetric configurations}\label{Sect1_general}
The action of the Einsteinian General Relativity in the presence of a minimally-coupled real SF  $\phi$ is\footnote{The metric signature is (+\,-\,-\,-), $R^\alpha_{\,\,\beta\gamma\delta}=-R^\alpha_{\,\,\beta\delta\gamma}=\partial_\gamma\Gamma^\alpha_{\beta\delta}-...\,;\,\,R_{\mu\nu}=R^\alpha_{\,\,\mu\alpha\nu}$.}
\begin{equation}\label{action}
S=\int  d^4\, x \sqrt{-g}\left(-\frac{R}{2\kappa}+\frac{1}{2}\partial_{\mu}\phi\partial^{\mu}\phi-V(\phi)\right)\,,
\end{equation}
where $\phi=\phi(t,r)$, $\kappa=8\pi G$ ($c=1$).

The energy-momentum of scalar field $\phi $ corresponding to (\ref{action}) is
\begin{equation} \label{Tmunu}
T_{\mu \nu}=\partial_\mu\phi \, \partial_\nu\phi -g_{\mu\nu} \left[\frac12  \partial_\alpha\phi  \,\partial^\alpha\phi-V(\phi)\right]\,.
\end{equation}
We work with the spherically symmetric space-time in the Schwarzshild-like (curvature) coordinates
 \begin{equation}
    ds^2=e^\nu dt^2-e^\lambda dr^2-r^2dO^2 \,,
    \label{SS-metric}
\end{equation}
where 
$ \nu =\nu (t, r), \lambda =\lambda (t, r), \phi =\phi (t, r)$, $dO^2=d \theta^2+ (\sin\theta)^2 d\varphi^2$, radial variable $r>0$.

The Einstein equations in case of  
 \eqref{Tmunu},\eqref{SS-metric}   are as follows \cite{Landau}: 
\begin{align}
&\frac{\partial}{\partial r} \left[ r \left(e^{-\lambda}-1\right) \right]=\nonumber  \\ & =-\frac{\kappa}{2} 
r^2 \left[ 
 e^{-\nu}\left(\frac{\partial\phi}{\partial t}\right)^2+ e^{-\lambda}\left(\frac{\partial\phi}{\partial r}\right)^2+  2V(\phi)\right] \,,
    \label{Ein_1} 
    \end{align}  
    \begin{align}
&{re^{-\lambda}\frac{\partial\nu}{\partial r}+e^{-\lambda}-1} = \nonumber\\& = \frac{\kappa}{2} r^2 \left[ e^{-\nu}\left(\frac{\partial\phi}{\partial t}\right)^2+ e^{-\lambda}\left(\frac{\partial\phi}{\partial r}\right)^2-  2V(\phi)\right]\,,
\label{Ein_2}
\end{align}  
  \begin{equation}
\frac{\partial \lambda }{\partial t}=\kappa r\frac{\partial \phi }{\partial t}\frac{\partial \phi }{\partial r}\,,  
\label{Ein_3}
\end{equation}  
The scalar field equation corresponding to action (\ref{action}) is
\begin{align}
    \label{eq_field}
     &\frac{\partial}{\partial t}\left[e^{(\lambda-\nu)/2}\frac{\partial \phi}{\partial t}\right] -
  \frac{1}{r^2}   \frac{\partial}{\partial r}\left[r^2 e^{(\nu-\lambda)/2}\frac{\partial \phi}{\partial r}\right]
     =\nonumber \\& -e^{(\nu+\lambda)/2}V'(\phi)\,,\quad V'\equiv dV/d\phi\,.
\end{align}
 Combination of (\ref{Ein_1}) and (\ref{Ein_2}) yields
\begin{align}
\frac{\partial}{\partial r} \left( \nu+\lambda\right) &= {\kappa r}
  \left[ e^{\lambda-\nu}
 \left(\frac{\partial\phi}{\partial t}\right)^2+  \left(\frac{\partial\phi}{\partial r}\right)^2 \right] \,,
    \label{Ein1+} \\[3pt] 	
\frac12\frac{\partial}{\partial r} \left( \nu-\lambda\right) &= \frac{e^\lambda-1}{r} - \kappa r  e^{\lambda}V(\phi)\,,
\label{Ein2+} 
\end{align}
These equations will be used to study perturbations against static spherically symmetric (SSS)  background.

In the static case, only  equations    (\ref{Ein1+}),(\ref{Ein2+})  and (\ref{eq_field}) are needed for $\nu=\nu_0(r)$, $\lambda=\lambda_0(r)$, $\phi=\phi_0(r)$:  
\begin{align}
 \frac{d}{d r} \left( \nu_0+\lambda_0\right) &=  {\kappa r} 
     \left(\frac{d\phi_0}{d r}\right)^2   \,,
    \label{Ein1+SSS} \\[3pt] 	
\frac12\frac{d}{d r} \left( \nu_0-\lambda_0\right) &= \frac{e^{\lambda_0}-1}{r} - \kappa r  e^{\lambda_0}V(\phi_0)\,,
\label{Ein2+SSS} 
\end{align}
\begin{equation}
    \label{eq_field_SSS}
       \frac{1}{r^2}   \frac{d}{d r}\left[r^2 e^{(\nu_0-\lambda_0)/2}\frac{d \phi_0}{d r}\right]
     = e^{(\nu_0+\lambda_0)/2}V'(\phi_0)\,.
\end{equation}
The qualitative properties of the SSS configurations with scalar field described by (\ref{Ein1+SSS}),(\ref{Ein2+SSS}),(\ref{eq_field_SSS}) have been studied  \cite{ZhdSt} for a wide family of the scalar field potentials, which includes the power-law case
 \begin{equation}
     V(\phi)=V_0 \phi^{2n}\,.      \label{power-law-pot}
\end{equation}
 We consider (\ref{power-law-pot}) with   $n>2$, which  corresponds to the Coulomb asymptotic of the field for large values of radial variable (see, e.g., \cite{ZhdSt}, Appendix 1).

When considering the SSS background solutions  $ \nu=\nu_0(r),\lambda=\lambda_0(r),\phi=\phi_0(r)$, we rely on articles \cite{ZhdSt,SZA2}. The   qualitative properties of them are common for a wide class of SF potentials \cite{ZhdSt} including (\ref{power-law-pot}). The static solution is represented by a monotonically decreasing $|\phi_0(r)|$ that cannot be equal to zero in any nontrivial case.  Near 
the origin, for $r\to 0$ we have asymptotics 
\begin{equation}\label{Y_at_the_center}
     r^2e^{\nu_0-\lambda_0}\to Y_f\,,
 \end{equation}
 where the constant $Y_f<\infty$ is strictly positive \cite{ZhdSt,SZA2}.
 Functions $\nu_0,\lambda_0, \phi_0$ have a logarithmic asymptotic near the origin $r=0$, where the naked singularity is present   \cite{ZhdSt,SZA2}.

The solution of the system (\ref{Ein1+SSS}), (\ref{Ein2+SSS}), (\ref{eq_field_SSS}) with the potential (\ref{power-law-pot}) is specified uniquely by the following asymptotic conditions \cite{SZA2}
\begin{equation}\label{asympt}
   \nu_0(r)=-\mu_0(r)=-\frac{r_g}{r}\,,\quad \phi_0(r)=\frac{Q}{r}\,,
\end{equation}
where $r_g=2GM$, $M$ is the background configuration mass.   Parameter $Q$ characterizes the strength of the scalar field far from the system; this parameter will be referred to as the ``scalar charge''. Below we used the length units such that $r_g=2$;  without limiting generality, we assumed $Q>0$.

We solved numerically the system   (\ref{Ein1+SSS}), (\ref{Ein2+SSS}), (\ref{eq_field_SSS}) with the potential (\ref{power-law-pot})  for  $n=3,4$ and $5$ for various $Q$. Following \cite{SZA2}, we performed backward numerical integration from sufficiently large $r=r_0$ to smaller values of the radial variable. The initial conditions at $r_0$ were set according to the asymptotics (\ref{asympt}) for $r_g/r_0\ll1$, $|Q|/r_0\ll1$. 

\begin{figure}[tbp]
\includegraphics[width=.4\textwidth]{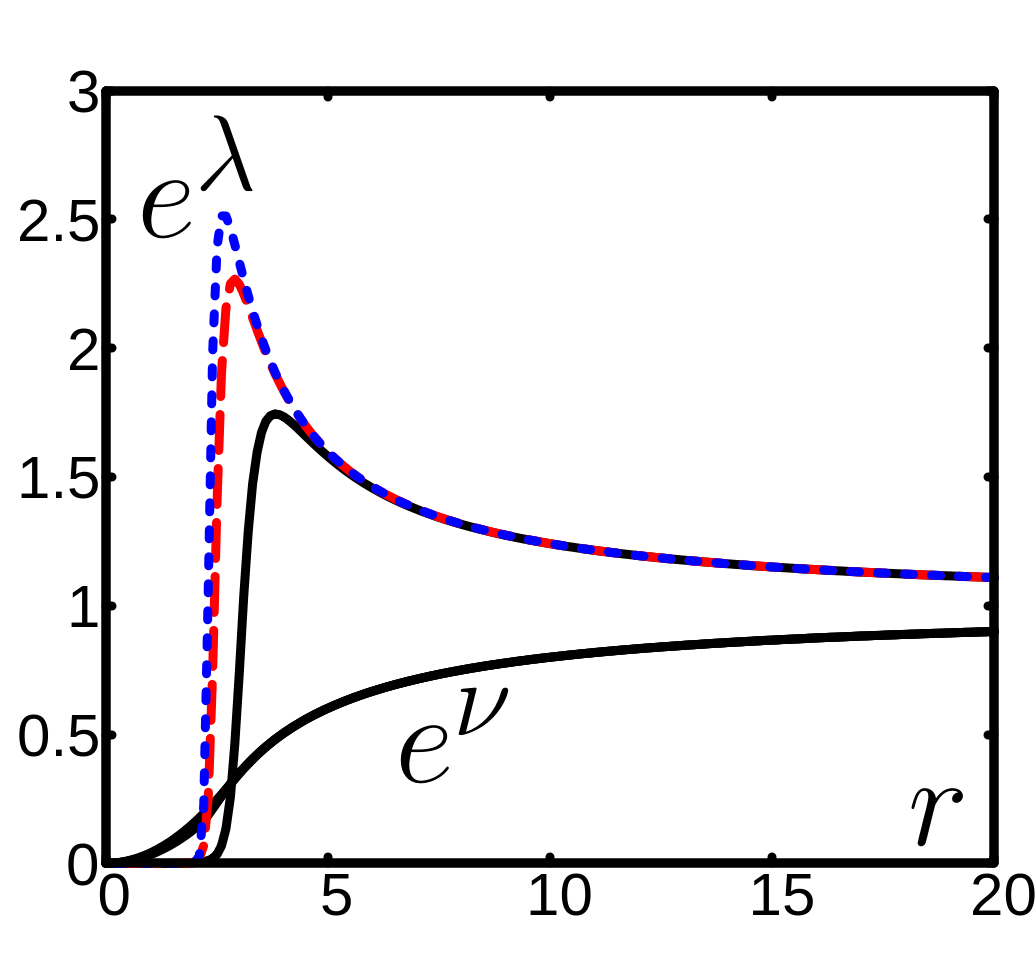}
     \caption{Typical dependencies of the background metric components upon the radial variable: $g_{tt}=e^{\lambda_0(r)}$ (lower curve, practically the same for all $n=3,4,5$) and $g_{rr}=e^{\nu_0(r)}$ (three curves with a maximum, top down: $n=5$ - blue short dashes, $n=4$ - red long dashes, $n=3$ - black solid line). In all the cases the scalar charge is $Q=1$, $r_g=2$. $V_0=1$.} 
\label{Fig1}  
\end{figure}

\begin{figure}[h!]
\includegraphics[width=.4\textwidth]{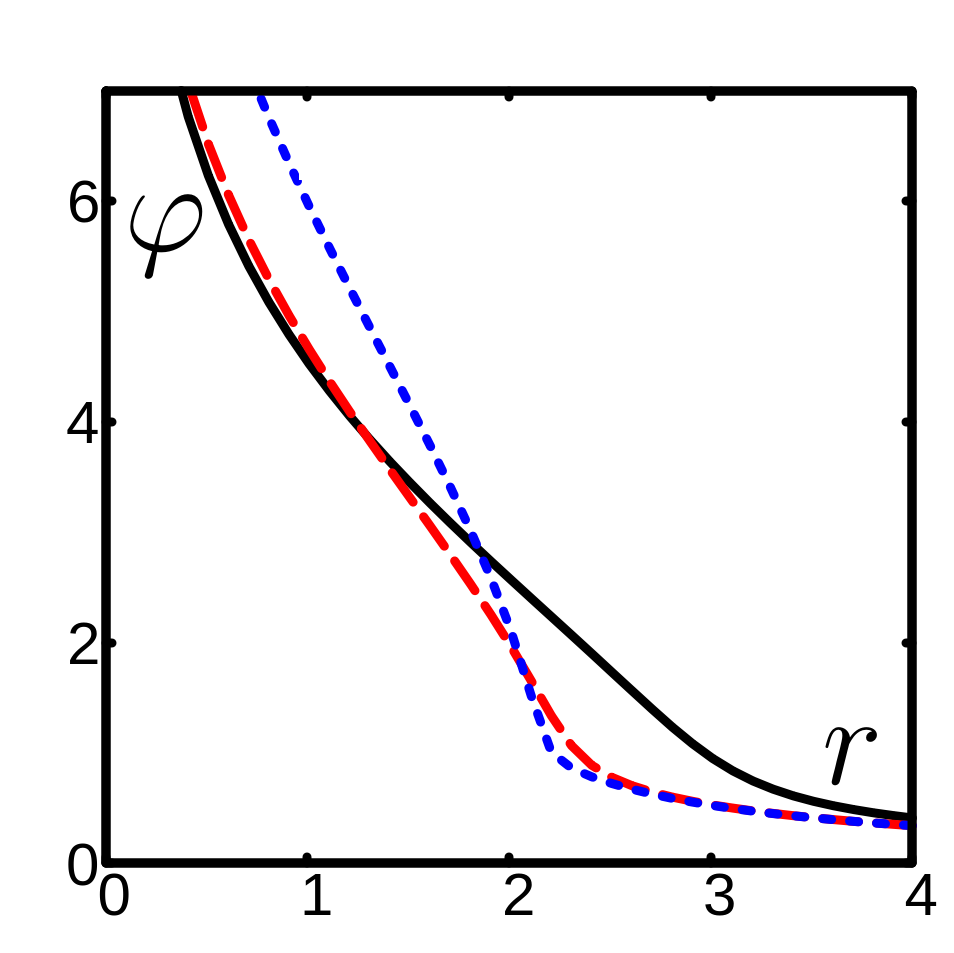}
     \caption{Typical dependencies of the background scalar field $\phi_0(r)$ for   $n=3$ (black solid line), $n=5$ (red long dashes), 
     $n=10$ (blue short dashes).
      In all the cases  $Q=1$, $r_g=2$. $V_0=1$.}
\label{Fig2}  
\end{figure}
Typical examples of solutions are shown in Figs. \ref{Fig1}, \ref{Fig2}. Numerical simulations show that the smaller $Q$, the higher maximum of $e^{\lambda_0(r)}$ near $r\simeq r_g$ and the closer to $r_g$ is the region, where the metric is approximately equal to the Schwarzschild one. 

\section{Radial perturbations}\label{Section2_perturbations}
Let us consider linearized stability against small radial perturbations of the static background by putting $\nu=\nu_0(r)+\delta \nu(t,r)$, $\lambda= \lambda_0(r) +\delta \lambda(t,r)$, $\phi= \phi_0(r)+\delta \phi(t,r)$. Given SSS solutions $ \nu_0(r),\lambda_0(r),\phi_0(r)$, our aim is to check the existence of divergent modes, i.e. solutions for $ \delta \nu(t,r),\delta \lambda(t,r),\delta \phi(t,r)$,  which satisfy appropriate boundary conditions but grow exponentially with time.

The procedure as a whole is completely similar  to that used in \cite{ZhdStSht2024} to study the SSS configurations in the $f(R)$ gravity. 
We shall express perturbations $ \delta \nu(t,r), \delta \lambda(t,r) ,\delta \phi(t,r)$  by means of one  function $\Phi =r  \delta \phi$ satisfying a second order ``master'' equation.

Linearized equation (\ref{Ein_3}) is
 \begin{equation*}
\frac{\partial\delta \lambda }{\partial t}=\kappa r\frac{\partial \delta\phi }{\partial t}\frac{d \phi_0 }{d r}\,,  
 \end{equation*}  
whence, omitting the static term, we have
\begin{equation}
   \delta \lambda  =\kappa r \delta \phi  \frac{d \phi_0  }{d r}\,.
\label{linEin_3-0++}
\end{equation}
Linearized equation  (\ref{Ein1+})  is 
\begin{equation}
\frac12 \frac{\partial}{\partial r} \left(\delta \nu+\delta \lambda \right)=\kappa r\frac{\partial \delta\phi}{\partial  r}\frac{d \phi_0 }{d r} \,, 
\label{delta+}
\end{equation}
whence using (\ref{linEin_3-0++}) we get
\begin{equation}
\frac12 \frac{\partial}{\partial r} \left(\delta \nu-\delta \lambda \right)=-\kappa\delta\phi \frac{d (r\phi_0) }{d r}\,. 
\label{delta-}
\end{equation}
This equation  can also be obtained directly from (\ref{Ein2+}) using expression for $V'$ following from (\ref{eq_field}); it will be used to obtain a linearized version of (\ref{eq_field}). 

Equations (\ref{delta+}) and  (\ref{delta-}) yield
\begin{equation}
\delta \nu=\kappa\int\limits_r^\infty dr' \left[\delta\phi(r') \frac{d (r'\phi_0(r')) }{d r'}-r'\frac{d \phi_0(r') }{d r'}\frac{\partial \delta\phi(r')}{\partial  r'}\right] \,, 
\label{delta-nu}
\end{equation}
where we take into account that $\nu(\infty)=0$. Thus, all perturbations are expressed by means of $\delta \phi$.

After linearization of equation (\ref{eq_field}), using (\ref{delta+}) and (\ref{delta-}), we obtain the master equation in terms of $\Phi =r  \delta \phi$: 
\begin{equation}
\frac{\partial ^2\Phi }{\partial t^2}-\frac{\partial ^2\Phi }{\partial r_*^2}+W_{eff}(r)\Phi =0\,,
\label{Eq_master_t}
\end{equation}
where $r_*$ is the ``tortoise'' radial coordinate 
$   {d r_* }/{d r} =e^{\frac{\lambda_0-\nu_0}{2}}\,, \quad r_*(0)=0$,  
and the effective potential is (cf. \cite{Clayton1998})
\begin{align}
    \label{Weff_1}
&W_{eff}(r)=   e^{\nu_0}V''(\phi_0 )+ \frac{e^{\nu_0-\lambda_0 }}{r}\frac{d}{dr}\frac{\nu_0-\lambda_0}{2}\nonumber+\\&+\kappa \phi_0' e^{\nu_0}\left[ rV'(\phi_0) +e^{-\lambda_0 }
\frac{d (r\phi'_0) }{d r} \right]   \,,
\end{align}
$\phi_0 '\equiv d\phi_0/dr$.

\begin{figure}[t]
\centering
\includegraphics[width=.4\textwidth]{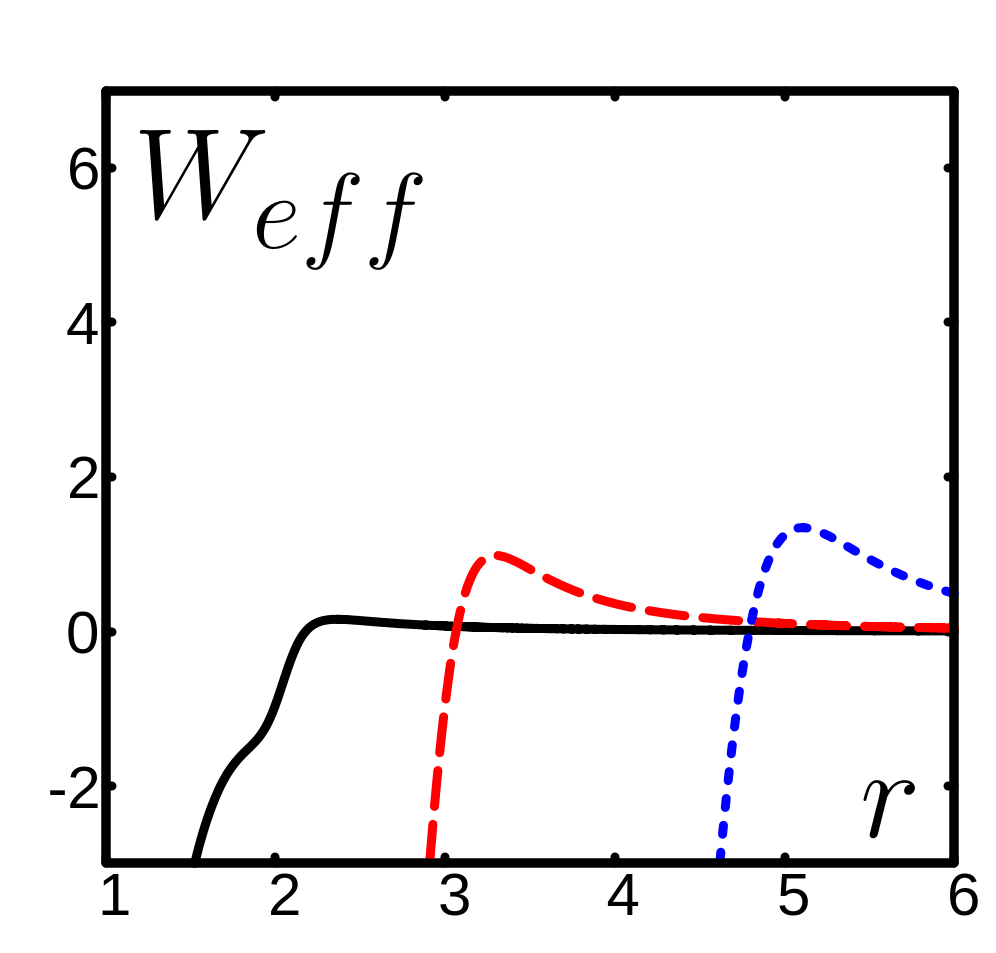}
     \caption{The effective potential $W_{eff}(r)$ for the perturbations in case of $n=3$ for the background configuration parameters:   $Q=1.5$ (blue short dashes), $Q=1$ (red long dashes), $Q=0.5$ (black solid line).$V_0=1$.} 
\label{Wef_n3}  
\end{figure}

\begin{figure}[t]
\centering
\includegraphics[width=.40\textwidth]{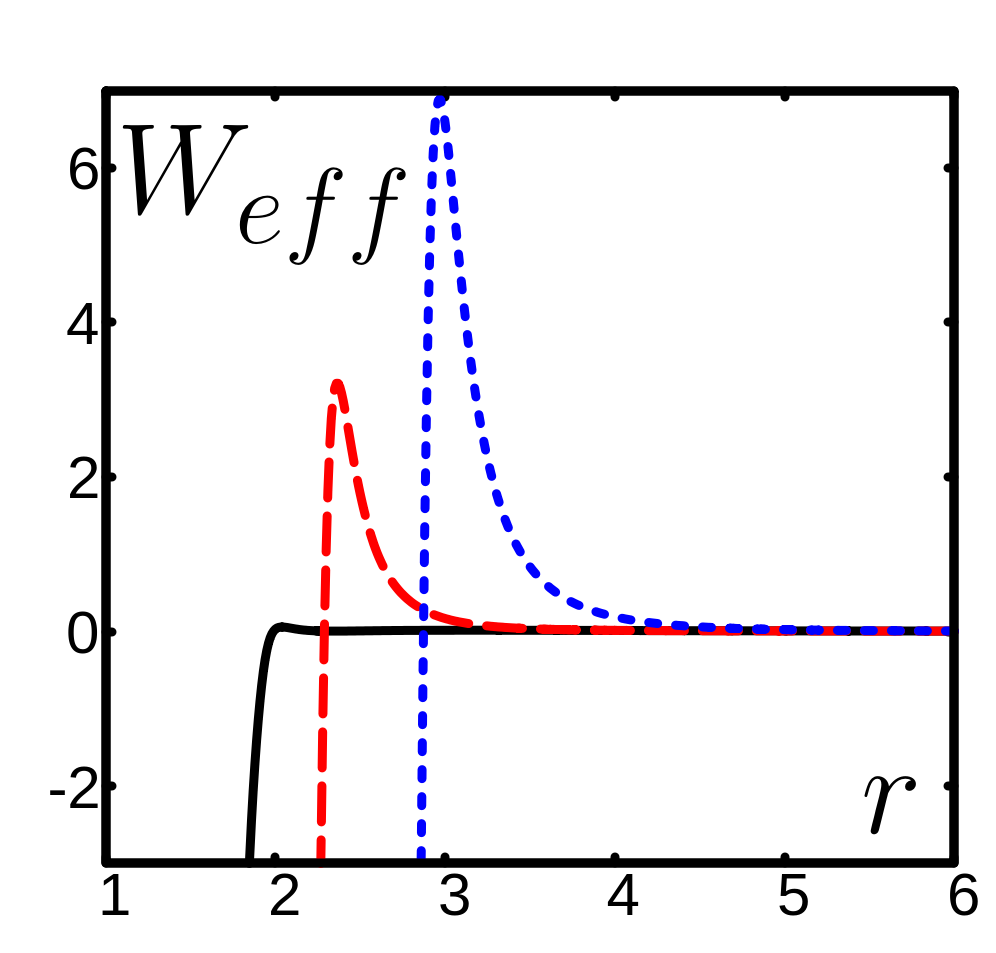}
\caption{The effective potential $W_{eff}$ for the same values of $Q$ as in Fig. \ref{Wef_n3}   in case  $n=5, \,V_0=1$.} 
\label{Wef_n5}  
\end{figure}

Taking into account equations (\ref{Ein2+SSS}) and (\ref{eq_field_SSS}) for static background we get the final expression:
\begin{align}
   & W_{eff}(r)\nonumber=-\frac{e^{\nu_0 -\lambda_0 }}{r^2}+ \frac{e^{\nu_0} }{r^2}
 \left[1-\kappa r^2V(\phi_0 )\right]
\left[1-{\kappa r^2} {\phi ^{'2}_0 } \right]+\\& +e^\nu_0 \left[ 2\kappa r  V'(\phi_0 )\phi'_0+ V''(\phi_0)\right] \,. \label{Weff}
\end{align}
We have $W_{eff}(r)\approx\ {r_g}/{r^3}$ for $r\to\infty$. Taking into account that $r^2e^{\nu_0-\lambda_0}\to const >0$ for $r\to 0$ \cite{ZhdSt}, we obtain that near the center the main input into $W_{eff}$ in (\ref{Weff}) is due to the first singular term. The examples of $W_{eff}(r)$ are shown in Figs. \ref{Wef_n3}, \ref{Wef_n5}.

Then we separate the dependence on time, putting $\Phi\sim e^{-i\omega t}$ that corresponds to the transition to Fourier-transformed quantities.  This yields
\begin{equation}
\frac{d^2\Phi }{d r_*^2}=(\Omega^2+W_{eff})\Phi \,,\quad \Omega^2=-\omega^2\,.
\label{Eq_master}
\end{equation}
If  there are solutions of (\ref{Eq_master}) with $\Omega >0$ (imaginary $\omega$) satisfying appropriate boundary conditions, then there exist solutions of (\ref{Eq_master_t}) $\Phi(t)\sim e^{\Omega t}$ and, therefore, we have instability. 

The well-posed problem requires the boundary conditions  at the center and at the infinity. 
We have $W_{eff}\sim r^{-4}$, and general asymptotic solution for $r\to 0$ is $\Phi(r)=C_1r+C_2r \ln r$.   We act in a usual way as in, e.g., \cite{Clayton1998,Gibbons2005,StSavZh2024} by imposing condition
\begin{equation}
\Phi(r)=C_1r\,,\quad d\Phi/dr=C_1\,, 
\label{boundary-condition-center}
\end{equation}
for $r\to 0$.  
This asymptotics is the only possible  to guarantee  regularity and smallness of the perturbation as $r\to 0$. Note that the choice of any $C_1\ne 0$ does not affect the asymptotics at infinity and we put for definiteness\footnote{In fact, $C_2$ and the solution for perturbation must be small. However, we deal with linear equation (\ref{Eq_master}) and its solution can be rescaled by multiplying by a sufficiently small factor without any affecting the result regarding stability. } $C_1=1$.   

For $r\to \infty$ we use standard requirement  that the perturbation must be square integrable, which gives the  asymptotics
\begin{equation}\label{boundary-condition-infty}
\Phi(r)\sim \exp(-\Omega r_*)\,,\,\Omega>0\,  .
\end{equation}

The numerical procedure to solve the problem (\ref{Eq_master}), (\ref{boundary-condition-center}), (\ref{boundary-condition-infty}) uses a ``shooting method'' that can be described as follows. We solve (\ref{Eq_master}) with the initial conditions at sufficiently small $r=r_1$ according to (\ref{boundary-condition-center}). Any solution of  (\ref{Eq_master}) has the following asymptotic form as $r\to\infty$
\begin{equation}\label{condition_infty}
 \Phi(r)=A(\Omega)e^{\Omega r_*}+B(\Omega)e^{-\Omega r_*} \,.    
\end{equation}
Therefore, we look for zeros of the function
\begin{equation*} 
  A(\Omega)=\lim\limits_{r\to \infty} e^{-\Omega r_*}\Phi(r) \,.    
\end{equation*}
The value of $\Omega$ such that
\begin{equation}\label{condition_infty}
 A(\Omega)=0\,, 
\end{equation}
implements the solution of the Sturm-Liouville problem  (\ref{Eq_master}) with boundary conditions  (\ref{boundary-condition-center}), (\ref{boundary-condition-infty}). 

\begin{figure}[t]
\centering
\includegraphics[width=.40\textwidth]{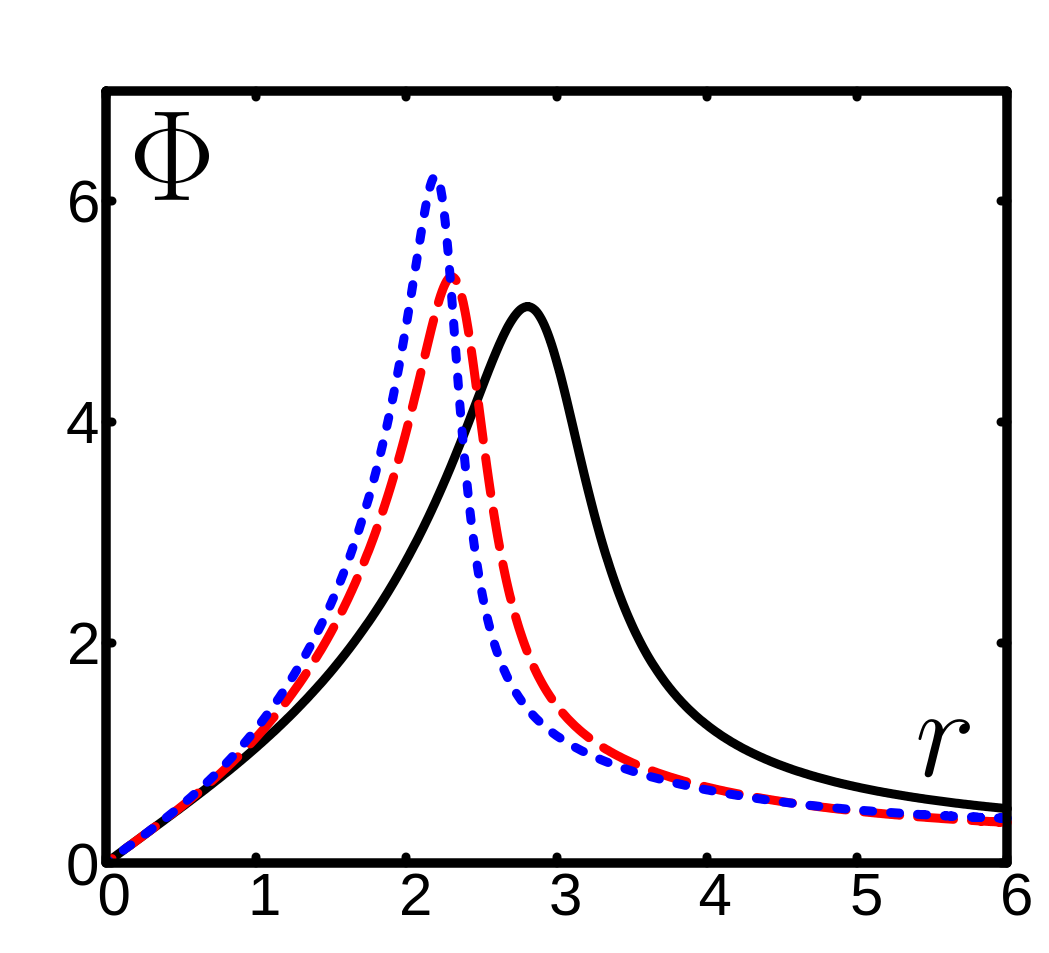}
     \caption{Typical examples of eigenfunctions $\Phi(r)$ in case of  $Q=1$ for the SF potentials with  $n=5$ (blue short dashes), $n=4$ (red long dashes), $n=3$ (black solid line); $V_0=1$. }
\label{eigenfunctions}  
\end{figure}

\begin{figure}[h!]
\centering
\includegraphics[width=.4\textwidth]{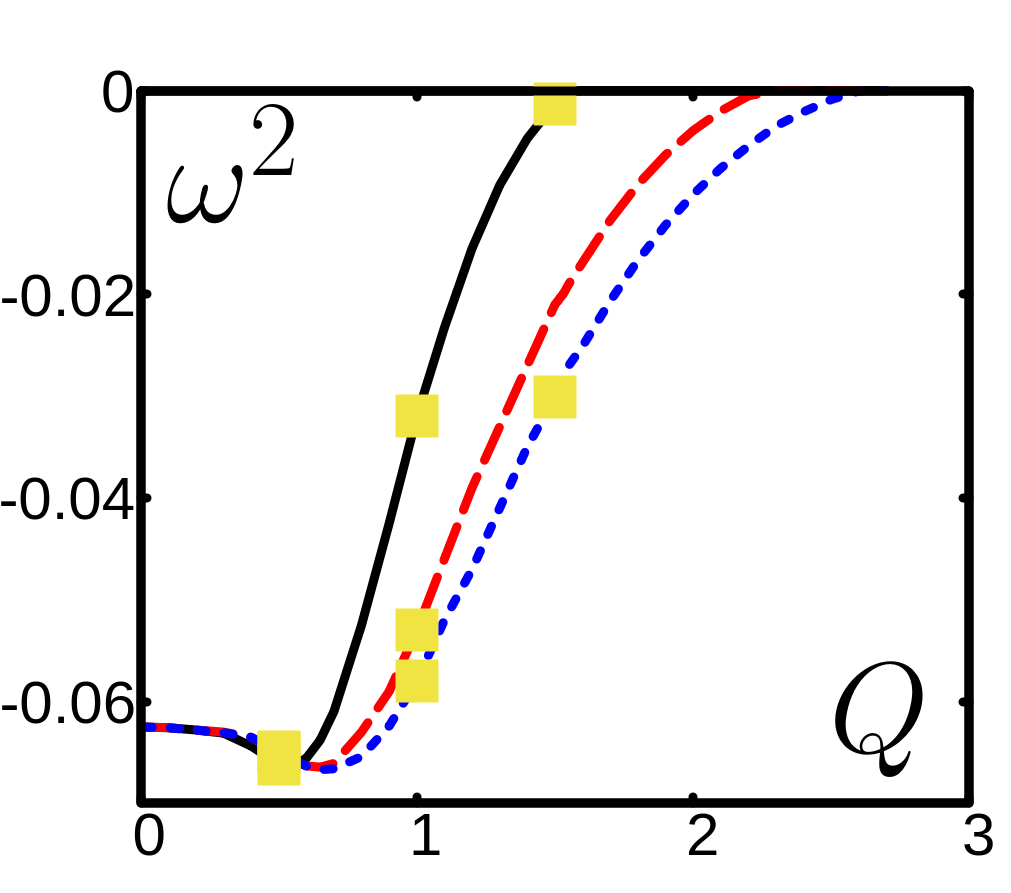}
\caption{Eigenvalues as a functions of $Q$ for the SF potentials with $n=5$ (blue short dashes), $n=4$ (red long dashes), $n=3$ (black solid line), same as in Fig. \ref{eigenfunctions} ($V_0=1$). The yellow squares for $Q=1$ show eigenvalues for perturbations presented in Fig. \ref{eigenfunctions} and to potentials  in Figs. \ref{Wef_n3}, \ref{Wef_n5} shown by red long dashed lines. Squares for $Q=0.5$  corresponds to solid lines in Figs. \ref{Wef_n3}, \ref{Wef_n5} and for $Q=1.5$ -- to short dashed lines.} 
\label{eigenvalues}  
\end{figure}

\begin{figure}[h!]
\centering
\includegraphics[width=.4\textwidth]{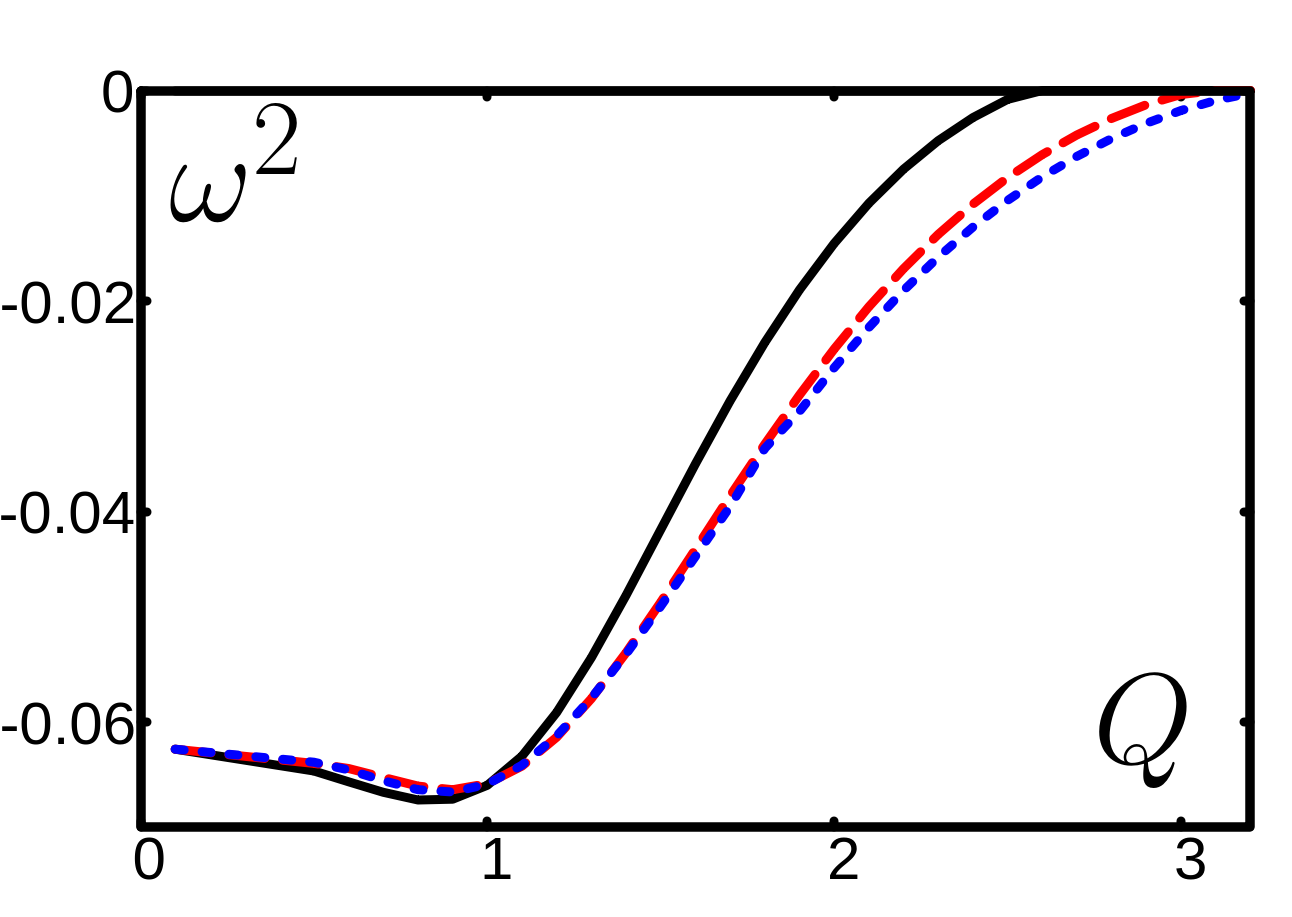}
\caption{Same as in Fig. \ref{eigenvalues}, but $V_0=0.1$. } 
\label{eigenvalues1}  
\end{figure}

 The unstable modes correspond to $\omega^2=-\Omega^2<0$. 
The examples of eigenfunctions $\Phi$ satisfying the boundary conditions are shown in Fig. \ref{eigenfunctions}. 
Fig. \ref{eigenvalues} shows the eigenvalues for $n=3,4,5$, $V_0=1$ and different $Q$. 

Denote  $Q_{max}(V_0, n)$ -- maximal value of $Q$, such that   SSS background configurations are unstable for $Q<Q_{max}$ for fixed $V_0, n$. For $Q>Q_{max}$,  unstable modes (with $\Omega>0$) and asymptotics (\ref{boundary-condition-infty})  do not exist.   Therefore, configurations with $Q>Q_{max}$  are linearly stable with respect to the monopole perturbations. 
In particular, for  $n=3$ at $Q_{max}\approx 1.53$, for $n=4$ at $Q_{max}\approx 2.15$ and for $n=5$ at $Q_{max}\approx 2.52$.

 \begin{figure}[h!]
\centering
\includegraphics[width=.4\textwidth]{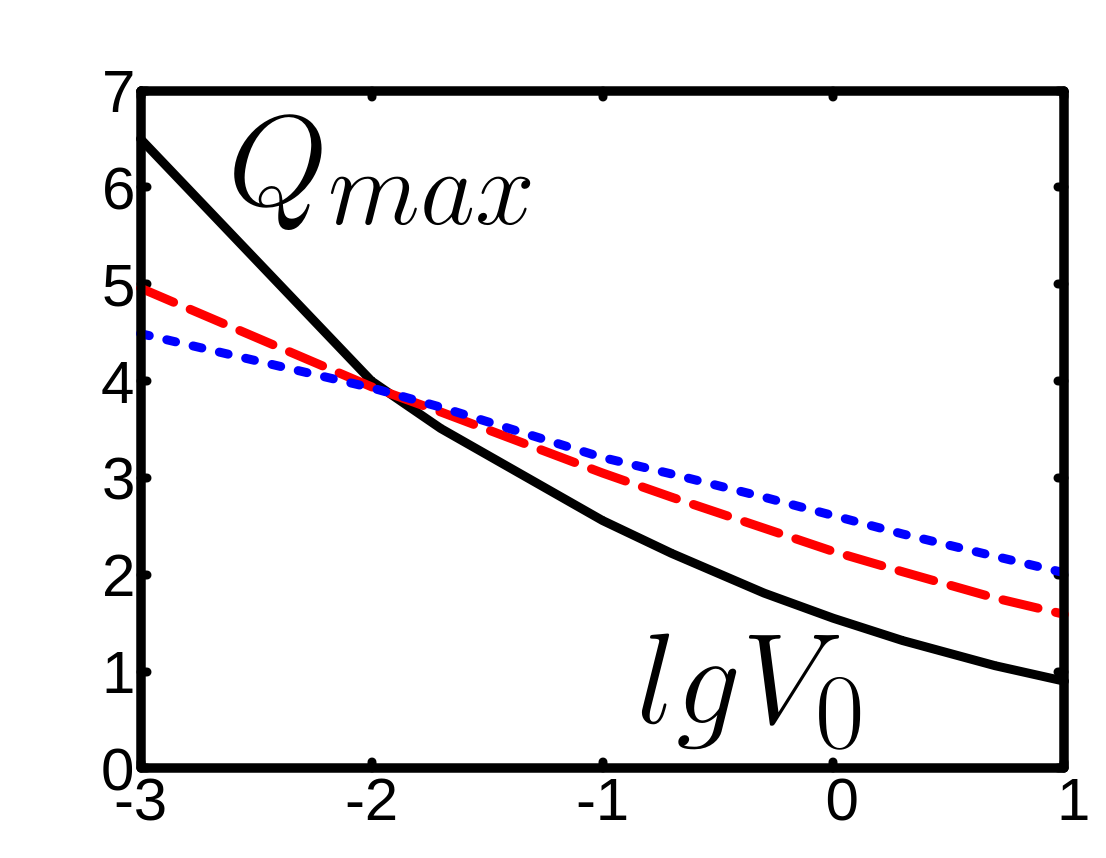}
\caption{Dependence $Q_{max}(V_0)$ for different power-law exponents in (\ref{power-law-pot}) $n=5$ (blue short dashes), $n=4$ (red long dashes), $n=3$ (black solid line). For each $n$, the unstable modes exist in the $Q-V_0$ plane below the corresponding curve and  disappear above it. It is interesting to note that there is a common intersection point  $Q_{max}=3.9$ for $V_0=0.015$ for all curves with different $n$. } 
\label{eigenvalues2}  
\end{figure}

\section{Conclusions\label{Section-conclusions}}
We studied asymptotically flat static spherically symmetric configurations with power law SF potentials (\ref{power-law-pot}) within General Relativity. The choice  $n=3,4,5$ in (\ref{power-law-pot}) ensures the Coulomb asymptotic behavior of the  field at spatial infinity. 
According to general results \cite{ZhdSt} such configuration has a naked singularity at the center.   Leaving aside the question of the origin of such configurations, we focused on the issues of linear stability of the SSS background. 

The equation for the perturbations  have been reduced to one-dimensional  master equation. The effective potential in this equation is singular at the origin, therefore it was important to impose a correct asymptotic condition  for perturbations at $r\to 0$. We acted in a standard way in analogy to \cite{Clayton1998,Gibbons2005,StSavZh2024}.  The corresponding condition (\ref{boundary-condition-center}) ensures  regularity of solutions and well-defined dynamics (cf., e.g., \cite{Wald1980,Ishibashi-Hosoya1999,Ishibashi-Wald2003}). 

By investigating linear radial perturbations,  we have shown that SSS solutions with  smaller values of the scalar charge $Q$  exhibit unstable modes characterized by exponentially growing perturbations. For example, in case of $V_0=1$ the unstable modes exist for   $Q<1.53$ if $n=3$,  $Q<2.15$ if $n=4$ and $Q<2.52$, if  $n=5$ (see Fig. \ref{eigenvalues} and analogous example in Fig.  \ref{eigenvalues1}). This confirms the expectation that such naked singularities are linearly unstable and thus unphysical, in accordance with the Penrose cosmic censorship conjecture \cite{Penrose1965, Penrose2002}.

On the other hand, for sufficiently large $Q$ and fixed $n$ the unstable modes disappear. 
This should not  be interpreted as a definitive  evidence of stability yet, but rather as a motivation for further analysis  involving  other types of perturbations  in future studies. It was demonstrated   \cite{StSavZh2024} that in a fairly general case axial perturbations lead to the wave equation with a positive effective potential showing linear stability of a background configuration. However, the stability against polar perturbations in the specific problem considered still needs a detailed analysis.

\vskip3mm \textit{Acknowledgements.}
A.V.T and V.I.Z.  acknowledge partial support from scientific program "Astronomy and Space Physics" of Taras Shevchenko National University of Kyiv. Work of V.I.Z. has been supported in part by National Research Foundation of Ukraine under project No. 2023.03/0149.

\newpage

%


\end{document}